\documentclass[onecolumn,twoside]{IEEEtran}
\usepackage{mathpazo}
\usepackage{times}
\usepackage{amsmath}
\usepackage{amsfonts}
\usepackage{latexsym}
\usepackage{amssymb}
\usepackage{mathrsfs}
\usepackage{upref}
\usepackage{theorem}
\usepackage{graphicx}
\usepackage{psfrag}
\usepackage{cite}
\usepackage{multirow}
\usepackage{color}

\theoremstyle{plain}
\theorembodyfont{\normalfont\slshape}

\newtheorem{thm}{Theorem$\!$}
\newenvironment{theorem}
{\begin{thm}\hspace*{-1ex}{\bf.}}{\end{thm}}

\newtheorem{lem}[thm]{Lemma$\!$}

\newtheorem{alg}[thm]{Algorithm$\!$}

\newtheorem{prop}[thm]{Proposition$\!$}

\newtheorem{cor}[thm]{Corollary$\!$}

\newtheorem{defn}[thm]{Definition$\!$}

\newtheorem{xmpl}[thm]{Example$\!$}
\newenvironment{example}{\begin{xmpl}\hspace*{-1ex}{\bf.}}{\hfill$\Box$\end{xmpl}}

\newtheorem{cnstr}{Construction$\!$}

\newcounter{enumrom}
\renewcommand{\theenumrom}{(\roman{enumrom})}

\makeatletter
\renewcommand{\@endtheorem}{\endtrivlist}
\makeatother

\makeatletter
\renewcommand{\thefigure}{{\@arabic\c@figure}}
\renewcommand{\fnum@figure}{{\bf Figure\,\thefigure}}
\makeatother

\newcommand{\pf}{{\bf Proof: }}

\newcommand{\be}[1]{\begin{equation}\label{#1}}
\newcommand{\ee}{\end{equation}}

\renewcommand{\leq}{\leqslant}

\renewcommand{\geq}{\geqslant}

\newcommand{\Cref}[1]{Co\-ro\-lla\-ry\,\ref{#1}}

\newcommand{\C}{\mbox{${\cal C}$}}

\newcommand{\lan}{\mbox{$\langle$}}
\newcommand{\ran}{\mbox{$\rangle$}}
\newcommand{\qed}{\hfill$\Box$\\[1ex]}

\newcommand{\uu}{\mbox{$\underline{u}$}}
\newcommand{\al}{\alpha}

\newcommand{\xor}{\oplus}

\newcommand{\ux}{\underline{x}}

\newcommand{\uy}{\underline{y}}
\newcommand{\ug}{\underline{g}}

\newcommand{\eq}{\mbox{$\,=\,$}}

\outer\def\proclaim #1. #2\par{\medbreak
 \noindent{\bf#1.\enspace}{\sl#2\par}%
 \ifdim\lastskip<\medskipamount \removelastskip\penalty55\medskip\fi}

\begin{document}


\title{\LARGE\bf On Generalized Expanded Blaum-Roth Codes}

\author{\large
Mario~Blaum \\
IBM Research Division-Almaden\\
650 Harry Road\\
San Jose, CA 95120, USA \\
Mario.Blaum@ibm.com, mblaum@hotmail.com}


\maketitle

\begin{abstract}
Expanded Blaum-Roth (EBR) codes consist of $n\times n$ arrays 
such that
lines of slopes $i$, $0\leq i\leq r-1$ for $2\leq r<n$, as well as vertical lines,
have even parity. The codes are MDS with respect to columns, i.e.,
they can recover any $r$ erased columns, if and only if $n$ is a
prime number. Recently a
generalization of EBR codes, called generalized expanded Blaum-Roth
(GEBR) codes, was presented. GEBR codes consist of $p\tau\times
(k+r)$ arrays, where $p$ is prime and $\tau\geq 1$, such that lines
of slopes $i$, $0\leq i\leq r-1$, have even parity and every column
in the array, when regarded as a polynomial, is a multiple of $1\xor x^{\tau}$.
In particular, it was shown that when
$p$ is an odd prime number, 2 is primitive in $GF(p)$ and $\tau\eq
p^j$, $j\geq 0$, the GEBR
code consisting of $p\tau\times (p-1)\tau$ arrays is MDS. We
extend this result further by 
proving that GEBR codes consisting of $p\tau\times p\tau$ arrays are MDS if and
only if $\tau\eq p^j$, where $0\leq j$ and $p$ is {\em any} odd prime.  
\end{abstract}


\section{Introduction}
\label{int}
Blaum-Roth (BR) codes~\cite{br} consist of $(n-1)\times n$ arrays such that,
assuming a zero row is appended to an array, the lines of slope $i$
(with a toroidal topology), $0\leq i\leq r-1$ for $2\leq r\leq n-1$ on the resulting
$n\times n$ array, have even parity. It was proven in~\cite{br} that
such a code is MDS (on columns) if and only if $n$ is a prime number. 

A related more recent construction is the family of
expanded Blaum-Roth (EBR) codes~\cite{bdh,bh}, consisting on $n\times
n$ arrays with even parity on lines of
slope $i$, $0\leq i\leq r-1$ for $2\leq r\leq n-1$, and in addition,
with even parity on columns (vertical lines). A column has even
parity if and only if, when considered as a polynomial, it is a
multiple of $1\xor x$.  EBR codes are also MDS if and
only if $n$ is a prime number.

EBR codes were recently extended in~\cite{wh}, where generalized
expanded Blaum-Roth (GEBR) codes are presented. In GEBR codes, 
the codewords are $n\times (k+r)$ arrays, where $n\eq p\tau$, $p$
an odd prime number, and the lines of slope $i$, $0\leq i\leq r-1$
for $2\leq i\leq r-1$, like in the
case of EBR codes, have even parity, while the columns, when considered as
polynomials, are multiples of $1\xor x^{\tau}$. The GEBR codes
in~\cite{wh} are denoted as $GEBR(n,p,k,r)$. Notice that EBR codes
are the special case of GEBR codes in which $\tau\eq 1$, i.e.,
$GEBR(p,p,k,r)$ with $k+r\leq p$.

In~\cite{wh} it was proven that if 2 is primitive in $GF(p)$, then the
code $GEBR(p^{j+1},p,k,r)$ with $k+r\leq p^j(p-1)$ is MDS, i.e., any
$r$ erased columns can be recovered from the remaining $k$ columns. 
We will extend this result to $GEBR(p^{j+1},p,k=p^{j+1}-r,r)$ codes
with no restrictions on the odd primes $p$, i.e., 2 may be
primitive on $GF(p)$ or not. Observe that neither BR or EBR codes had
restrictions on $p$, so our result generalizes completely the 
construction of EBR codes. 
We will assume without loss of
generality that the arrays are
square, i.e., $n\times n$ arrays, since
when $n\times (k+r)$ arrays with $k+r<n$ are desired, we can simply pad
$n-k-r$ columns with zeros and then ignore such columns when writing
the arrays (the process known as shortening of a code~\cite{ms}).  

Section~\ref{cons} gives the construction of the GEBR codes while
section~\ref{MDS} presents necessary and sufficient conditions for a 
GEBR code to be MDS. We illustrate both the construction and the
properties of the codes with examples. 

We end the paper by drawing some conclusions.

\section{Construction of GEBR codes}
\label{cons}
Let $n$ be a positive integer. Given an integer $m$, let $\lan
m\ran_n$ be the unique integer $j$, $0\leq j\leq n-1$, such that
$j\equiv m\;(\bmod\;n)$. For example, $\lan 8\ran_6\eq 2$.

The following definition of a GEBR code is similar to the one
given in~\cite{wh}.

\begin{defn}
\label{def1}
{\em
Let $n\eq p\tau$, where $p$ is a prime number, and
let $r$ be an integer such that $2\leq
r<n$. Then, a code $GEBR(n,p,r)$ over a field $GF(q)$ of
characteristic 2 consists of the $n\times n$ arrays
$(a_{u,v})_{_{0\leq v\leq n-1}^{0\leq u\leq n-1}}$ 
such that, if $\ux\eq (x_0,x_1,\ldots,x_{n-1})$ is any column of such
an array, 

\begin{eqnarray}
\label{eq1}
\bigoplus_{\ell=0}^{p-1}\,x_{\ell\tau +u}&=&0\quad {\rm
for}\quad 0\leq u\leq \tau-1
\end{eqnarray}
and

\begin{eqnarray}
\label{eq2}
\bigoplus_{j=0}^{n-1}\,a_{\lan u-\ell j\ran_n,j}&=&0\quad {\rm
for}\quad 0\leq \ell\leq r-1\quad {\rm and}\quad 0\leq u\leq n-1.
\end{eqnarray}

}
\end{defn}

In Definition~\ref{def1}, (\ref{eq1}) denotes $\tau$ parities that
each column in the array must satisfy, while (\ref{eq2}) states that
the arrays must satisfy parities along all the lines of slope $\ell$
(with a toroidal topology), where $0\leq \ell\leq r-1$.
It is easy to see that condition~(\ref{eq1}) on a vector of length
$n\eq p\tau$ is equivalent to requiring that the vector, when viewed as a
polynomial on $\al$, is a multiple of $1\xor \al^{\tau}$~\cite{wh},
i.e., if $\ux\eq (x_0,x_1,\ldots,x_{n-1})$ and
$\ux(\al)\eq\bigoplus_{i=0}^{n-1}\,x_i\al^i$, $\ux$
satisfies~(\ref{eq1}) if and only if

\begin{eqnarray}
\label{eq40}
\ux(\al)&=&\ux'(\al)(1\xor \al^{\tau})\;\;{\rm
where}\;\;\deg(\ux'_i(\al))\leq n-\tau -1.
\end{eqnarray}

We illustrate
Definition~\ref{def1} in the following examples. 

\begin{example}
\label{ex1}
{\em
Let $n\eq p$, $p$ a prime number, in Definition~\ref{def1}, i.e., the
code is a $GEBR(p,p,r)$ code. Then
$\tau\eq 1$ and (\ref{eq1}) becomes 

\begin{eqnarray}
\label{eq3}
\bigoplus_{\ell=0}^{p-1}\,a_{\ell,j}&=&0\quad {\rm
for}\quad 0\leq j\leq p-1.
\end{eqnarray}

Equation~(\ref{eq3}) means that in addition to even parity over
lines of slope $\ell$, $0\leq \ell\leq r-1$, vertical lines also have
even parity, or equivalently, that every column, when considered as a
polynomial, is a multiple of $1\xor x$. This one is the special case
of EBR codes, which were 
proven to be MDS~\cite{bdh,bh}. 
}
\end{example}

\begin{example}
\label{ex2}
{\em
Consider the $GEBR(6,3,2)$ code in
Definition~\ref{def1}. The following $6\times 6$
array is in $GEBR(6,3,2)$:

$$
\begin{array}{|c|c|c|c|c|c|}
\hline
0&1&0&1&0&0\\\hline
1&0&1&1&1&0\\\hline
1&0&1&0&0&0\\\hline
0&0&1&1&1&1\\\hline
1&1&1&1&0&0\\\hline
1&0&0&0&0&1\\\hline
\end{array}
$$

We notice that this code is not MDS. In effect, assume that columns 0
and 3 are erased:

$$
\begin{array}{|c|c|c|c|c|c|}
\hline
&1&0&&0&0\\\hline
&0&1&&1&0\\\hline
&0&1&&0&0\\\hline
&0&1&&1&1\\\hline
&1&1&&0&0\\\hline
&0&0&&0&1\\\hline
\end{array}
$$

Then, the following array gives an alternative solution:

$$
\begin{array}{|c|c|c|c|c|c|}
\hline
1&1&0&0&0&0\\\hline
0&0&1&0&1&0\\\hline
1&0&1&0&0&0\\\hline
1&0&1&0&1&1\\\hline
0&1&1&0&0&0\\\hline
1&0&0&0&0&1\\\hline
\end{array}
$$

Since there is more than one solution, the code is not MDS. The
necessary and sufficient conditions for a GEBR code to be MDS to be
given in Theorem~\ref{theo1} explain the reason.
}
\end{example}

\begin{example}
\label{ex3}
{\em
Consider now the $GEBR(6,2,2)$ code
according to Definition~\ref{def1}. Now $\tau\eq 3$. 
The following $6\times 6$
array is in $GEBR(6,2,2)$:

$$
\begin{array}{|c|c|c|c|c|c|}
\hline
0&1&1&1&1&0\\\hline
1&1&1&0&0&1\\\hline
1&0&1&1&1&0\\\hline
0&1&1&1&1&0\\\hline
1&1&1&0&0&1\\\hline
1&0&1&1&1&0\\\hline
\end{array}
$$
We notice that this code is not MDS either. In effect, assume that columns 0
and 3 are erased:

$$
\begin{array}{|c|c|c|c|c|c|}
\hline
\phantom{0}&1&1&\phantom{0}&1&0\\\hline
&1&1&&0&1\\\hline
&0&1&&1&0\\\hline
&1&1&&1&0\\\hline
&1&1&&0&1\\\hline
&0&1&&1&0\\\hline
\end{array}
$$
Then, the following array gives an alternative solution:

$$
\begin{array}{|c|c|c|c|c|c|}
\hline
1&1&1&0&1&0\\\hline
0&1&1&1&0&1\\\hline
0&0&1&0&1&0\\\hline
1&1&1&0&1&0\\\hline
0&1&1&1&0&1\\\hline
0&0&1&0&1&0\\\hline
\end{array}
$$
}
\end{example}

\begin{example}
\label{ex30}
{\em
Consider the $GEBR(9,3,3)$ code
according to Definition~\ref{def1}. Now $\tau\eq 3$. 
The following $9\times 9$
array is in\\ $GEBR(9,3,3)$:

$$
\begin{array}{|c|c|c|c|c|c|c|c|c|}
\hline
0&0&1&1&0&0&1&0&1\\\hline
0 &1 &0 &0 &1 &1 &0 &0 &1\\\hline
0 &1 &1 &0 &1 &1 &1 &1 &0\\\hline
0 &1 &1 &1 &1 &1 &1 &1 &1\\\hline
1 &0 &0 &0 &1 &0 &0 &1 &1\\\hline
1 &0 &0 &0 &0 &1 &1 &1 &0\\\hline
0 &1 &0 &0 &1 &1 &0 &1 &0\\\hline
1 &1 &0 &0 &0 &1 &0 &1 &0\\\hline
1 &1 &1 &0 &1 &0 &0 &0 &0\\\hline
\end{array}
$$

We can see that each column satisfies condition~(\ref{eq1}), while every
horizontal line and every line of slope 1 or 2 has even parity. This
code is MDS, as we will see in Theorem~\ref{theo1}. Let us point out
that with the construction in~\cite{wh}, an MDS code over $9\times 6$
arrays was obtained. 
}
\end{example}

\section{Necessary and sufficient conditions for GEBR codes to be
MDS} 
\label{MDS}
Given $n\eq p\tau$, consider the ring of polynomials modulo $1+x^n$ and let
$\al^i$ denote a rotation of a vector of length $n$ $i$ times to the
right. Hence, $\al^n\eq 1$. We will use both vectors of length $n$
and their polynomial representations in $\al$ to describe 
polynomials in $GF(2)[x]$ modulo $1\xor x^n$. If $\ux$ denotes a vector,
$\ux(\al)$ denotes the corresponding polynomial in $\al$. For
example, for $n\eq 5$, if $\ux\eq (1,0,1,1,0)$ then
$\ux(\al)\eq 1\xor\al^2\xor\al^3$. We also denote the zero
polynomial simply by~0. 

An equivalent way of describing a $GEBR(n,p,r)$ code as given by Definition~\ref{def1}
is through the $r\times n$ Reed-Solomon type of parity check matrix~\cite{bdh,bh,wh} 

\begin{eqnarray}
\label{eq4}
H(n,r)&=&\left(
\begin{array}{ccccc}
1&1&1&\ldots &1\\
1&\al &\al^2&\ldots &\al^{n-1}\\
1&\al^2 &\al^4&\ldots &\al^{2(n-1)}\\
\vdots &\vdots &\vdots &\ddots &\vdots \\
1&\al^{r-1} &\al^{2(r-1)}&\ldots &\al^{(n-1)(r-1)}\\
\end{array}
\right).
\end{eqnarray}

Hence, $X(\al)\eq (\ux_0(\al),\ux_1(\al),\ldots,\ux_{n-1}(\al))\in
GEBR(n,p,r)$ if and only if each $\ux_i(\al)$ satisfies~(\ref{eq40})
and 


\begin{eqnarray}
\label{eq41}
H(n,r)X(\al)^{\rm T}&=&(\overbrace{0,0,\ldots,0}^r)^{\rm T}.
\end{eqnarray}

We can see, as stated above, that~(\ref{eq40}) corresponds
to~(\ref{eq1}), while~(\ref{eq41}) corresponds to~(\ref{eq2}), where $H(n,r)$ 
is given by~(\ref{eq4}).
From~(\ref{eq4}) and~(\ref{eq41}), since solving for $r$ erasures involves
``inverting'' an $r\times r$ Vandermonde matrix, a $GEBR(n,p,r)$ code
will be MDS if and only, for any $i$ such that $1\leq i\leq n-1$ and
for any polynomial modulo $1\xor x^n$ $\uu(\al)$
satisfying~(\ref{eq40}), the equation

\begin{eqnarray}
\label{eq5}
(1\xor\al^i)\ux(\al)\eq\uu(\al)
\end{eqnarray}
has a unique solution $\ux(\al)\eq\uy(\al)$
satisfying~(\ref{eq40}). Moreover, without
loss of generality, we may 
assume that $\uu(\al)\eq 0$. When $p$ is an odd prime number and $\tau\eq 1$,
recursion~(\ref{eq5}) with
$\uu(\al)\eq 0$ has the unique solution $\ux(\al)\eq 0$: it is the
especial case of EBR codes~\cite{bdh,bh}.

The necessary and sufficient conditions for a $GEBR(n,p,r)$ code to
be MDS are given by the following theorem: 

\begin{theorem}
\label{theo1}
{\em
Let $\C$ be a $GEBR(n,p,r)$ code, $n\eq p\tau$, according to Definition~\ref{def1}.
Then $\C$ is MDS if and only $p$ is an odd prime and $\tau\eq p^j$,
where $j\geq 0$.
}
\end{theorem}

\noindent\pf As stated above, code $\C$ is MDS if and only (\ref{eq5}) has a unique
solution $\ux(\al)\eq 0$ when $\uu(\al)\eq 0$ for $1\leq i\leq n-1$.

Assume first that $p\eq 2$, hence, $n\eq 2\tau$, and
$\tau>1$ ($\tau\eq 1$ is a trivial case). 
Let $i\eq \tau$. We claim, $(1\xor\al^{\tau})\ux(\al)\eq 0$ has a 
solution $\ux(\al)\neq 0$ satisfying~(\ref{eq40}), and hence $GEBR(2\tau,2,r)$ is
not MDS. In effect, let $\ux(\al)\eq 1\xor\al^{\tau}$. In particular,
$\ux(\al)$ trivially satisfies~(\ref{eq40}) and
$(1\xor\al^{\tau})\ux(\al)\eq (1\xor\al^{\tau})^2\eq 1\xor\al^{2\tau}\eq 0$
since $\al^{2\tau}\eq\al^n\eq 1$.

Next let $p>2$ and $\tau\eq p^jm$ (equivalently, $n\eq p^{j+1}m$)
with $j\geq 0$ and $\gcd(p,m)\eq 1$. Assume first that $m>1$. Then, let

\begin{eqnarray}
\label{eq6}
\ux_0(\al)&=&
1\xor\al^{p^{j+1}}\xor\al^{2p^{j+1}}\xor\ldots\xor\al^{(m-1)p^{j+1}}\\ 
\label{eq7}
\ux_1(\al)&=&\al^{p^j}\xor\al^{p^j+p^{j+1}}\xor\al^{p^j+2p^{j+1}}\xor\ldots\xor\al^{p^j+(m-1)p^{j+1}} 
\end{eqnarray}
and $\ux(\al)\eq\ux_0(\al)\xor\ux_1(\al)$. Notice that
$\ux_1(\al)\eq\al^{p^j}\ux_0(\al)$ and also

\begin{eqnarray*}
\al^{p^{j+1}}\ux_0(\al)&=&
\al^{p^{j+1}}\xor\al^{2p^{j+1}}\xor\al^{3p^{j+1}}\xor\ldots\xor\al^{mp^{j+1}}\\ 
&=&\al^{p^{j+1}}\xor\al^{2p^{j+1}}\xor\al^{3p^{j+1}}
\xor\ldots\xor 1\quad =\quad \ux_0(\al)
\end{eqnarray*}
since $\al^{mp^{j+1}}\eq\al^n\eq 1$. Similarly,
$\al^{p^{j+1}}\ux_1(\al)\eq\ux_1(\al)$ and hence $(1\xor
\al^{p^{j+1}})\ux(\al)\eq (1\xor
\al^{p^{j+1}})(\ux_0(\al)\xor\ux_1(\al))\eq 0$. Then, $\ux(\al)\neq 0$ is a 
solution to $(1\xor\al^{p^{j+1}})\ux(\al)\eq 0$. It
remains to be shown that 
$\ux(\al)$ satisfies~(\ref{eq40}) in order to prove that the code is not MDS.

From (\ref{eq6}), the set of powers of $\al$ in $\ux_0(\al)$ given by

\begin{eqnarray*}
G&=&\{0,p^{j+1},2p^{j+1},\ldots ,(m-1)p^{j+1}\} 
\end{eqnarray*}
is a subgroup of size $m$ of
the additive group $Z_n$ of integers modulo $n\eq p^{j+1}m$. In
effect, since
$\gcd(p,m)\eq 1$, 

\begin{eqnarray}
\label{eq8}
up^{j+1}&\not\equiv&vp^{j+1}\;(\bmod\;\tau)\;\;{\rm for}\;\;0\leq u<v\leq m-1.
\end{eqnarray}
Also, 

\begin{eqnarray*}
p^j+G&=&\{p^j,p^j+p^{j+1},p^j+2p^{j+1},\ldots ,p^j+(m-1)p^{j+1}\}
\end{eqnarray*}
is a coset of $G$, which by~(\ref{eq7}), corresponds to the powers of
$\al$ in $\ux_1(\al)$.
Since $G$ and $p^j+G$ are disjoint, $\ux(\al)$
has weight $2m$. Since $\gcd(p,m)\eq 1$, there is an $\ell$,
$0\leq\ell\leq m-1$, such that $p\ell\equiv -1\;(\bmod\;m)$.
Consider the assignment $f\,:\,G\rightarrow p^j+G$

\begin{eqnarray}
\label{eq9}
f(ip^{j+1})&\eq &p^j+\lan \ell+i\ran_mp^{j+1}\;\;{\rm
for}\;\;0\leq i\leq m-1.
\end{eqnarray}
It is clear that $f$ is onto, and hence also 1-1. 

Notice that, since $p\ell\equiv -1\;(\bmod\;m)$,
$1+p\ell\eq um$ for some $u$,
so, for any $i$, $0\leq i\leq m-1$ 
\begin{eqnarray*}
1+\lan\ell+i\ran_mp-ip&=&u'm \;\;{\rm for }\;\;{\rm some}\;\;u',
\end{eqnarray*}
and multiplying both sides by $p^j$, since $p^jm\eq \tau$,
by~(\ref{eq9}), we obtain,

\begin{eqnarray*}
p^j+\lan\ell+i\ran_mp^{j+1}-ip^{j+1}\;\;=\;\;f(ip^{j+1})-ip^{j+1}&=&u'mp^j\;\;\eq \;\;u'\tau \;\;{\rm for }\;\;{\rm some}\;\;u',
\end{eqnarray*}
or equivalently, 

\begin{eqnarray}
\label{eq10}
f(ip^{j+1})-ip^{j+1}&\equiv&0\;(\bmod\;\tau)\;\;{\rm for}\;\;0\leq
i\leq m-1.
\end{eqnarray}

Consider $\ug_i(\al)\eq\al^{f(ip^{j+1})}\xor \al^{ip^{j+1}}$. By~(\ref{eq10}),
$\ug_i(\al)$ is divisible by $1\xor\al^{\tau}$, hence $\ug_i(\al)$
satisfies~(\ref{eq40}). Grouping the non-zero elements of $\ux(\al)$ in
pairs, we obtain 

\begin{eqnarray*}
\ux(\al)&=& \bigoplus_{i=0}^{m-1}\,\ug_i(\al).
\end{eqnarray*}

Since each $\ug_i(\al)$ is divisible by $1\xor\al^{\tau}$, then also
$\ux(\al)$ is divisible by $1\xor\al^{\tau}$, satisfying~(\ref{eq40}).

Finally, assume that $m\eq 1$, then $\tau\eq p^j$ and $n\eq p^{j+1}$.
If $\gcd(i,p)\eq 1$, then the only solution to
$(1\xor\al^i)\ux(\al)\eq 0$ is $\ux(\al)\eq 
0$~\cite{wh}. Otherwise $i\eq up^s$, where $\gcd(u,p)\eq 1$, $1\leq
u\leq p^{j+1-s}-1$ and $1\leq s\leq j$. Assume that
$\ux(\al)$ satisfies~(\ref{eq40}), $\ux(\al)\neq 0$ and
$(1\xor\al^{up^s})\ux(\al)\eq 0$. Without 
loss of generality, let 
$\ux(\al)\eq\bigoplus_{v=0}^{n-1}x_v\al^v$ and $x_0\eq
1$. Since $x_0\xor x_{up^s}\eq 0$, then $x_{up^s}\eq 1$. Since 
$x_{\lan (\ell-1) up^s\ran_n}\xor x_{\lan\ell up^s\ran_n}\eq 0$ for
$1\leq \ell\leq p^{j+1-s}-1$, by induction, 
$x_{\lan\ell up^s\ran_n}\eq 1$ for $0\leq \ell\leq p^{j+1-s}-1$.
Since $\gcd(u,p)\eq 1$, the sets $\{\lan\ell up^s\ran_n\;:\;0\leq\ell\leq p^{j+1-s}-1\}$
and $\{\ell p^s\;:\;0\leq\ell\leq p^{j+1-s}-1\}$ coincide, hence,
$x_{\ell p^s}\eq 1$ for $0\leq \ell\leq p^{j+1-s}-1$. In particular,
taking $\ell\eq vp^{j-s}$ for $0\leq v\leq p-1$,
$x_{vp^j}\eq x_{v\tau}\eq 1$, so
$\bigoplus_{v=0}^{p-1}x_{v\tau}\eq 1$, contradicting condition~(\ref{eq1}).
\qed

We illustrate the proof of Theorem~\ref{theo1} in the next examples.

\begin{example}
\label{ex4}
{\em
Consider the $GEBR(6,3,r)$ code according to Definition~\ref{def1} with
$2\leq r\leq 5$.
Following the notation of Theorem~\ref{theo1}, $\tau\eq m\eq 2$ and 
$j\eq 0$. According to Theorem~\ref{theo1}, the $GEBR(6,3,r)$ code is
not MDS. In effect, according to~(\ref{eq6}) and~(\ref{eq7}),

\begin{eqnarray*}
\ux_0(\al)&=&1\xor\al^{3}\\ 
\ux_1(\al)&=&\al\xor\al^{4}
\end{eqnarray*}
and $\ux(\al)\eq\ux_0(\al)\xor\ux_1(\al)\eq 1\xor\al\xor\al^3\xor\al^4$. 
We can see that
$(1\xor\al^3)\ux(\al)\eq 0$, so this recursion has a non-zero
solution. Following the proof of Theorem~\ref{theo1}, notice that
$\ux(\al)\eq (1\xor\al^4)\xor (\al\xor \al^3)$. Each of these binomials is
divisible by $1\xor\al^2$, so $\ux(\al)$ satisfies~(\ref{eq40})
and the code $GEBR(6,3,r)$ is not MDS.
}
\end{example}

\begin{example}
\label{ex5}
{\em
Consider the $GEBR(45,3,r)$ code according to Definition~\ref{def1}.
Following the notation of Theorem~\ref{theo1}, $\tau\eq 15$,
$m\eq 5$ and $j\eq 1$. According to Theorem~\ref{theo1}, the $GEBR(45,3,2)$ code is
not MDS. By~(\ref{eq6}) and~(\ref{eq7}),

\begin{eqnarray*}
\ux_0(\al)&=&
1\xor\al^{9}\xor\al^{18}\xor\al^{27}\xor\al^{36}\\ 
\ux_1(\al)&=&\al^3\xor\al^{12}\xor\al^{21}\xor\al^{30}\xor\al^{39}\\
\ux(\al)\eq\ux_0(\al)\xor\ux_1(\al)&=&
1\xor\al^3\xor\al^{9}\xor\al^{12}\xor\al^{18}\xor\al^{21}\xor\al^{27}\xor\al^{30}\xor\al^{36}
\xor\al^{39}
\end{eqnarray*}
We can verify that $(1\xor\al^9)\ux(\al)\eq 0$, giving a non-zero
solution to the recursion. Following the proof of
Theorem~\ref{theo1}, we have $G\eq\{0,9,18,27,36\}$ and
$3+G\eq\{3,12,21,30,39\}$. Let $\ell\eq 3$, then
$3\ell\equiv\;-1(\bmod\;5)$. By the assignment
$f:G\rightarrow 3+G$ given by~(\ref{eq9}), we have $f(0)\eq 30$, $f(9)\eq 39$,
$f(18)\eq 3$, $f(27)\eq 12$ and $f(36)\eq 21$. Grouping together the
pairs of binomials $\al^{9i}\xor \al^{f(9i)}$ for $0\leq i\leq 4$, we have
$$\ux(\al)\eq
(1\xor\al^{30})\xor(\al^9\xor\al^{39})\xor(\al^3\xor\al^{18})\xor(\al^{12}\xor\al^{27})\xor(\al^{21}\xor\al^{36}).$$ 
 We can see that each of the binomials is divisible by
 $1\xor\al^{\tau}\eq 1\xor\al^{15}$,
 so $\ux(\al)$ satisfies~(\ref{eq40}).


A different code consisting of $45\times 45$ arrays we may consider
is a $GEBR(45,5,r)$ code. In this case, $\tau\eq m\eq 9$ and $j\eq 0$.
By~(\ref{eq6}) and~(\ref{eq7}), 

\begin{eqnarray*}
\ux_0(\al)&=&\bigoplus_{u=0}^8\al^{5u}\\
\ux_1(\al)&=&\bigoplus_{u=0}^8\al^{1+5u}\\ 
\end{eqnarray*}

Making $\ux(\al)\eq\ux_0(\al)\xor\ux_1(\al)$,
$(1\xor\al^5)\ux(\al)\eq 0$ gives a non-zero solution to the recursion.
Following the proof of
Theorem~\ref{theo1}, we have $G\eq\{0,5,10,15,20,25,30,35,40\}$ and
$1+G\eq\{1,6,11,16,21,26,31,36,41\}$. Let $\ell\eq 7$, then
$5\ell\equiv\;-1(\bmod\;9)$. By the assignment
$f:G\rightarrow 1+G$ given by~(\ref{eq9}), we have $f(5i)\eq \lan 36+
5i\ran_{45}$ for $0\leq i\leq 8$.
Grouping together the
pairs of binomials $\al^{5i}\xor \al^{f(5i)}$, we have
$\ux(\al)\eq \bigoplus_{i=0}^8 (\al^{5i}\xor\al^{\lan 36+5i\ran_{45}})$.
 We can see that each of the binomials is divisible by $1\xor\al^{\tau}\eq 1\xor\al^9$,
 so $\ux(\al)$ satisfies~(\ref{eq40}).
}
\end{example}

\begin{example}
\label{ex6}
{\em
Consider the $GEBR(27,3,r)$ code according to Definition~\ref{def1}.
Following the notation of Theorem~\ref{theo1}, $\tau\eq 9$,
$m\eq 1$ and $j\eq 2$. According to Theorem~\ref{theo1}, the $GEBR(27,3,r)$ code is
MDS. This means, $(1\xor\al^i)\ux(\al)\eq 0$ has the unique solution
$\ux(\al)\eq 0$ for $1\leq i\leq 26$. This is certainly true for
$\gcd(3,i)\eq 1$. Assume that $i\eq
3$. If $x_0\eq 1$, since $(1\xor\al^3)\ux(\al)\eq 0$, then,
$x_{3u}\eq 1$ for $0\leq u\leq 8$. In particular, $x_9\eq x_{18}\eq
1$, and hence $x_0\xor x_{\tau}\xor x_{2\tau}\eq x_0\xor x_{9}\xor
x_{18}\eq 1$, contradicting~(\ref{eq1}). 

The same result is obtained by taking as $i$ any multiple of 3. 
}
\end{example}

\section{Conclusions}
We have presented necessary and sufficient conditions for Generalized
Expanded Blaum-Roth codes, as defined in~\cite{wh}, to be MDS. The
encoding and decoding depend on efficient solving of recursions, as
described in~\cite{wh}, where also methods like the LU factorization
of Vandermonde matrices is given for very fast decoding. We refer the
reader to~\cite{wh} for the description of encoding and decoding
algorithms.

\end{document}